\documentclass[10pt,conference]{IEEEtran}

\usepackage[T1]{fontenc}

\usepackage{cite}
\usepackage{amsmath,amsthm,mathtools}

\usepackage{amsfonts,amssymb,amsbsy} 
\usepackage{esint} 

\usepackage{bm} 

\usepackage{array} 

\usepackage[caption=false,font=footnotesize]{subfig} 

\usepackage{stfloats} 

\usepackage{url} 

\usepackage{balance} 

\usepackage{dsfont}
\usepackage{textcomp}

\usepackage[multiple]{footmisc}

\usepackage{tikz}
\tikzset{every picture/.style={font issue=\footnotesize},
	font issue/.style={execute at begin picture={#1\selectfont}}
}
\usepackage{pgfplots}
\pgfplotsset{compat=newest}
\usetikzlibrary{backgrounds}
\usetikzlibrary{plotmarks}
\usetikzlibrary{positioning}
\usetikzlibrary{arrows.meta}
\usetikzlibrary{calc} 

\usepackage{xr-hyper}

\makeatletter
\newcommand*{\addFileDependency}[1]{
  \typeout{(#1)}
  \@addtofilelist{#1}
  \IfFileExists{#1}{}{\typeout{No file #1.}}
}
\makeatother

\usepackage[capitalize]{cleveref}
	\Crefname{figure}{Fig.}{Fig.}
	\Crefname{section}{Sec.}{Sec.}
	\Crefname{subsection}{Sec.}{Sec.}
	\Crefname{prop}{Proposition}{Proposition}
	\Crefname{lemma}{Lemma}{Lemma}
	\Crefname{equation}{}{}
	\Crefname{footnote}{Footnote}{Footnote}

\input{macros}

\renewcommand{\d}{{\bf d}}
\newcommand{\cd}{c\hspace{0.3mm}}
\newcommand{\NObs}{M}

  \newcommand{\AssAsyn}{} 
  
  \newcommand{\NoAssAsyn}{_{\text{N/A}}} 

  \newcommand{\MLE}{{}^\text{\scriptsize{\,MLE}}}
  \newcommand{\LSE}{{}^\text{\scriptsize{\,LSE}}}
  \newcommand{\MVUE}{{}^\text{\scriptsize{\,MVUE}}}
  \newcommand{\Hypo}[1]{\tilde{#1}} 
    
  \newcommand{\AssViaTau}{_{\hspace{.15mm}\text{by}\hspace{.3mm}\tau}} 
  \newcommand{\AssViaTauSync}{_{\hspace{.15mm}\text{by}\hspace{.3mm}\tau,\,\text{sync}}} 
  \newcommand{\AssViaDiff}{_{\hspace{.15mm}\text{by}\hspace{.3mm}\Delta}}
  \newcommand{\AssViaDiffPWA}{_{\hspace{.15mm}\text{by}\hspace{.3mm}\Delta,\,\text{PWA}}}

  \newcommand{\tauRMS}{\sigma_\tau }
  \newcommand{\errClock}{\epsilon}
  \newcommand{\errClockA}[1][]{\errClock\AnnotateNode{A}_{#1}}
  \newcommand{\errClockB}[1][]{\errClock\AnnotateNode{B}_{#1}}
  \newcommand{\errClockAEstimate}[1][]{{\hat\errClock}\AnnotateNode{A}_{#1}}
  
  \newcommand{\errMeasSymbol}{n}
  \newcommand{\errMeas}[1][]{\vect[#1]{\errMeasSymbol}}
  \newcommand{\errMeasA}[1][]{\errMeas[#1]\AnnotateNode{A}}
  \newcommand{\errMeasB}[1][]{\errMeas[#1]\AnnotateNode{B}}
  
  \newcommand{\AnnotateNode}[1]{^{\textnormal{\resizebox{3.3mm}{!}{\hspace{.2mm}(\hspace{.1mm}#1\hspace{.1mm})}}}}
  \newcommand{\AnnotateNodeLOWER}[1]{^{\resizebox{2.8mm}{!}{\hspace{.2mm}(\hspace{.1mm}#1\hspace{.1mm})}}}
  
  \newcommand{\delayTrueA}[1][]{\vect[#1]{\bar\tau}\AnnotateNode{A} }
  \newcommand{\delayTrueB}[1][]{\vect[#1]{\bar\tau}\AnnotateNode{B} }
  \newcommand{\delayMeasA}[1][]{\vect[#1]{\tau}\AnnotateNode{A} }
  \newcommand{\delayMeasB}[1][]{\vect[#1]{\tau}\AnnotateNode{B} }
    
  \newcommand{\AnnotateSorted}[1]{\check{#1}}
    
     \newcommand{\cirA}{h\AnnotateNode{A}}
     \newcommand{\cirB}{h\AnnotateNode{B}}
	 
 	\newcommand{\dirVectLetter}{e}
  	\newcommand{\dirMtxLetter}{E}
	\newcommand{\dirVectObsLetter}{f}
  	\newcommand{\dirMtxObsLetter}{F}
	
    \newcommand{\E}{\mathbf{\dirMtxLetter}} 
    
    \newcommand{\dirVect}[1][]{\vect{e}_{#1} }
    \newcommand{\dirVectA}[1][]{ 
      \ifthenelse{\isempty{#1}}
      {\vect{\dirMtxLetter}\AnnotateNode{A}}  
      {\vect{\dirVectLetter}\AnnotateNode{A}_{#1}}
     } 
    \newcommand{\dirVectB}[1][]{ 
      \ifthenelse{\isempty{#1}}
      {\vect{\dirMtxLetter}\AnnotateNode{B}}
      {\vect{\dirVectLetter}\AnnotateNode{B}_{#1} }
    } 
    \newcommand{\dirVectObs}[1][]{ 
      \ifthenelse{\isempty{#1}}
       {\vect{\dirMtxObsLetter}}
       {\vect{\dirVectObsLetter}_{#1}}  
    }
    \newcommand{\dirVectObsA}[1][]{
      \ifthenelse{\isempty{#1}}
      {\vect{\dirMtxObsLetter}\AnnotateNode{A}}
      {\vect{\dirVectObsLetter}\AnnotateNode{A}_{#1} }
    }
    \newcommand{\dirVectObsB}[1][]{
      \ifthenelse{\isempty{#1}}
      {\vect{\dirMtxObsLetter}\AnnotateNode{B}}
      {\vect{\dirVectObsLetter}\AnnotateNode{B}_{#1} }
    }
  
    \newcommand{\dirVectBSort}[1][]{
      \ifthenelse{\isempty{#1}}
      {\vect{\AnnotateSorted{\dirMtxLetter}}\AnnotateNode{B}}
      {\vect{\AnnotateSorted{\dirVectLetter}}\AnnotateNode{B}_{#1} }
    }
    \newcommand{\dirVectObsBSort}[1][]{
      \ifthenelse{\isempty{#1}}
      {\vect{\AnnotateSorted{\dirMtxObsLetter}}\AnnotateNode{B}}
      {\vect{\AnnotateSorted{\dirVectObsLetter}}\AnnotateNode{B}_{#1}}
    }
    
\newcommand{\delayDiff}[1][]{\vect[#1]{\Delta} }
\newcommand{\delayDiffTrue}[1][]{\vect[#1]{\bar\Delta} }

\newcommand{\delayDiffPerm}[2]{
    \ifthenelse{ \(\isempty{#1} \or \isempty{#2}\) \and \not\( \isempty{#1} \and \isempty{#2} \) } 
      {\error{}}
      {\vect[#1,#2]{\Delta}^{\text{perm}} }   
    }
    
  \newcommand{\posA}{\vect{p}\AnnotateNode{A} } 
  \newcommand{\posB}{\vect{p}\AnnotateNode{B} } 

\newcommand{\permSmybol}{\pi}
\newcommand{\perm}{\permSmybol} 

\newcommand{\permSetSymbol}{\Pi}
\newcommand{\permSet}[1][]{\ifthenelse{\isempty{#1}}{\permSetSymbol}{\permSetSymbol_{#1}}}

\newcommand\iid{\overset{\textnormal{iid}}{\sim}}

\newcommand\OurPaperTitle{%
Pairwise Distance and Position Estimators From Differences in UWB Channels to Observers}

\newcommand\OurIndexTerms{ultra-wideband ranging, distance estimation, relative localization, indoor localization, synchronization}

\begin{document}

\title{\OurPaperTitle}
\author{
\IEEEauthorblockN{Gregor Dumphart$^*$, Robin Kramer$^*$, and Armin Wittneben} 
\IEEEauthorblockA{\textit{Wireless Communications Group, D-ITET, ETH Zurich, Switzerland}\\
Email: \{dumphart, kramer, wittneben\}@nari.ee.ethz.ch\\
$^*\,$The authors contributed equally to this work%
}}

\maketitle
	
	\begin{abstract}
    We consider the problem of obtaining relative location information between two wireless nodes from the differences in their ultra-wideband (UWB) channels to observer nodes. Our approach focuses on the delays of multipath components (MPCs) extracted from the observed channels. For the two different cases of known and unknown MPC association between these channels, we present estimators for the distance and for the relative position vector between the two nodes. The position estimators require both MPC directions and MPC delays as input. All presented estimators exhibit very desirable technological properties: they do not require line-of-sight conditions, precise synchronization, or knowledge about the observer locations or about the environment. These advantages could enable low-cost wireless network localization in dynamic multipath environments. The exposition is complemented by a numerical evaluation of the estimation accuracy using random sampling, where especially the position estimators show the potential for great accuracy.
%
%
	\end{abstract}
	

	\section{Introduction}
	\label{sec:Intro}
	Wireless localization is a key requirement for many mobile applications in the Internet of Things, robotics, and social distance monitoring for pandemic contact tracing \cite{Hatke2020}. Many such applications concern dense and dynamic propagation environments, characterized by time-variant channels with rich multipath propagation and frequent line-of-sight (LOS) obstruction \cite{WitrisalSPM2016}. This poses a great challenge to accurate and reliable wireless localization and ranging.
For example, distance estimates from the received signal strength (RSS) tend to have large relative error due to large RSS fluctuations \cite{SchultenVTC2019}.
Time of arrival (TOA) distance estimates often have a substantial bias due to LOS obstruction, multipath, and synchronization problems \cite{DardariPIEEE2009,YuPLANS2020}. This causes large relative errors at short distances.
Trilateration of such inaccurate distance estimates results in inaccurate position estimates; ensuring enough anchors in LOS to all relevant mobile positions is often infeasible \cite{WitrisalSPM2016}.
Location fingerprinting is also not an all-round alternative for accurate localization: the training data is quickly rendered obsolete by time-varying environments \cite{HePC2016}.

State-of-the-art localization systems deal with these problems with methods such as soft information processing \cite{MazuelasTSP2018} and temporal filtering \cite{BuehrerPIEEE2018,LeitingerTWC2019}.
Various recent work \cite{WitrisalSPM2016,LeitingerTWC2019,YuPLANS2020} considers multipath as opportunity rather than interference: multipath-assisted UWB localization allows for improved accuracy and robustness if knowledge about the propagation environment is either available a-priori \cite{WitrisalSPM2016} or obtained with mapping \cite{LeitingerTWC2019}.
Also promising is the use of MPC direction information such as the angle of arrival (AOA), which can be measured robustly with millimeter-wave massive-MIMO systems \cite{BuehrerPIEEE2018,WenTWC2020}.
Further improvements are possible with cooperative (a.k.a. collaborative) network localization \cite{BuehrerPIEEE2018}. 



\begin{figure}[!ht]
\centering
\vspace{-1mm}
\subfloat[conventional paradigm]{
\begin{tabular}{c}
\ \ \ \ \ \
\includegraphics[height=15mm,trim=0 -7mm 0 0]{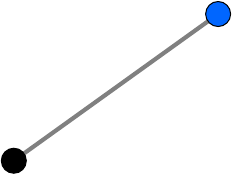}
\ \ \ \ \ \
\put(-55,31){$d$, $h(\tau)$}
\put(-49,-.5){\footnotesize{node A}}
\put(-12,46){\footnotesize\textcolor[rgb]{0,.2,.8}{node B}}
\vspace{-1mm}
\end{tabular}
\label{fig:Concept_Conventional}} \ \ \ 
\subfloat[proposed paradigm]{
\centering
\begin{tabular}{c}
\includegraphics[height=15mm,trim=0 -7mm -3mm 0]{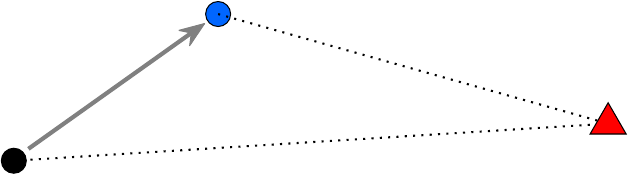}
\put(-109,30){$\d$}
\put(-75,3){$\cirA_o(\tau)$}
\put(-130,-.5){\footnotesize{node A}}
\put(-53,34){\textcolor[rgb]{0,.2,.8}{$\cirB_o(\tau)$}}
\put(-93,46){\footnotesize\textcolor[rgb]{0,.2,.8}{node B}}
\put(-30,4){\footnotesize{\textcolor[rgb]{.8,0,0}{observer $o$}}}
\vspace{-1mm}
\end{tabular}
\label{fig:Concept_Proposed}}
\ifdefined\SingleColumnDraft\else
\\
\fi
\subfloat[proposed signal processing]{
\centering
\raisebox{1.5mm}{
\resizebox{86mm}{!}{\begingroup
\begin{tikzpicture}[%
box/.style={draw,text width=9mm,minimum height=7mm,align=center,line width=1pt}]
\node[box,minimum height=16mm,minimum width=12mm,text width=10mm,fill=white] (AssocBox) {\baselineskip=10pt associate MPCs \par};
\node[box,right=4mm of AssocBox,minimum height=16mm,minimum width=15mm,text width=14mm,fill=white] (CompBox) {\baselineskip=10pt  estimation rule based on MPC differences \par}; 
\node[right=-1mm of AssocBox.144] (MpcOutA) {};
\node[right=-1mm of AssocBox.216] (MpcOutB) {};
\node[box,left=3mm of MpcOutA,fill=white](MpcA) {\baselineskip=10pt extract MPCs \par};
\node[box,left=3mm of MpcOutB,fill=white](MpcB) {\baselineskip=10pt extract MPCs \par};
\node[right=3mm of AssocBox.33] (AssocOutA) {};
\node[right=3mm of AssocBox.327] (AssocOutB) {};
\draw[-{Latex[length=2.5mm,width=1.6mm]},line width=1pt] (AssocBox.33) -- (AssocOutA.center);
\draw[-{Latex[length=2.5mm,width=1.6mm]},line width=1pt] (AssocBox.327) -- (AssocOutB.center);
\node[left=6.5mm of MpcA] (CirsA) {$\big\{\, \cirA_o(\tau) \,\big\}$};
\node[left=6.5mm of MpcB] (CirsB) {$\big\{\, \cirB_o(\tau) \,\big\}$};
\node[right=6.5mm of CompBox] (MyOutput) {$\hat d$ or $\hat\d$};
\begin{scope}[on background layer]
\fill[gray,thick,dotted,fill={rgb:black,1;white,4}] ($(MpcA.north west)+(-0.13,0.13)$)  rectangle ($(CompBox.south east)+(0.13,-0.13)$);
\end{scope}
\draw[-{Latex[length=2.5mm,width=1.6mm]},line width=1pt] (CirsA) -- (MpcA.180);
\draw[-{Latex[length=2.5mm,width=1.6mm]},line width=1pt] (CirsB) -- (MpcB.180);
\draw[-{Latex[length=2.5mm,width=1.6mm]},line width=1pt] (MpcA.0) -- (MpcOutA.center);
\draw[-{Latex[length=2.5mm,width=1.6mm]},line width=1pt] (MpcB.0) -- (MpcOutB.center);
\draw[-{Latex[length=2.5mm,width=1.6mm]},line width=1pt] (CompBox) -- (MyOutput);
\end{tikzpicture}
\endgroup}
}
\label{fig:SigProcBlackBox}}
\caption{Conventional versus proposed paradigm for relative localization of two wireless nodes.}
\label{fig:IntroConcepts}
\end{figure}
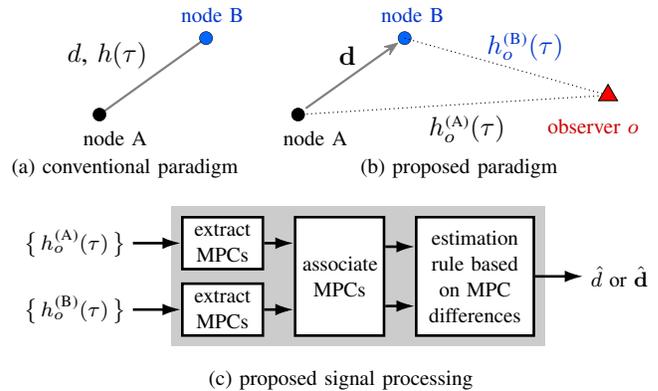

Our previous work \cite{DumphartICC2019} proposed an \textit{alternative paradigm} for pairwise localization between two nodes, with the goal of alleviating the outlined problems of wireless localization and ranging systems. It abandons the conventional notion that a distance estimate between two nodes A and B should be based on a direct measurement such as the TOA or RSS between them (\Cref{fig:Concept_Conventional}). 
Instead, the presence of one or more other nodes, henceforth called \textit{observers}, is considered (\Cref{fig:Concept_Proposed}). The paradigm relies on measurements of the channel impulse response (CIR) $\cirA_o(\tau)$ of the channel between node A and the observer as well as CIR $\cirB_o(\tau)$ between node B and the observer.
These CIRs can be obtained via channel estimation at the observer after transmitting training sequences at A and B, or vice versa.
In the case of UWB operation, the CIRs are descriptive signatures of the multipath environment \cite{WitrisalSPM2016}. 
We note that the CIRs $\cirA_o(\tau)$ and $\cirB_o(\tau)$ are similar for small distances $d$ but \textit{differ increasingly and systematically} with increasing $d$.
A good metric for CIR dissimilarity could give rise to an accurate estimate of $d$ or even of the relative position vector $\d$,
with the prospect of particularly good accuracy at short distances and no requirements for LOS connections.

Several approaches could be eligible for the realization of such an estimation rule, e.g., machine learning.
Our previous work \cite{DumphartICC2019} utilized parameters of extracted MPCs (\Cref{fig:SigProcBlackBox}), in particular the differences of MPC delays.
Based thereon we derived distance estimators under the assumption of known MPC association; specifically the maximum-likelihood estimate (MLE) and the minimum-variance unbiased estimate (MVUE) under propagation assumptions representative of indoor environments. The approach can be viewed as a profound propagation-geometric comparison of a measurement $\cirB(\tau)$ to a fingerprint $\cirA(\tau)$, although without any required preceding training.
For this approach we identified very desirable technological properties: it does not require line-of-sight conditions, knowledge about the observer locations or the environment, or precise synchronization.
To the best of our knowledge, this approach and the considered paradigm are not addressed by any existing work other than \cite{DumphartICC2019}.

%
In this paper we make the following novel contributions:
\begin{itemize}
\item
We state the distance MLE for the case of unknown MPC association, random delay measurement errors, and random MPC directions (uniform distribution in 3D).
\item
Assuming that the MPC directions are observable, we state the least-squares estimate (LSE) of the relative position vector for different cases: based on the observed delay differences or based directly on the observed delays.
The former promises higher robustness.
\item
We append a tailored scheme for estimating the MPC association from the MPC delays and directions.
\item
Based on a numerical evaluation of the estimation accuracy (using random sampling of MPC parameters), we describe the estimators' strengths and weaknesses.
\end{itemize}

All estimators are designed for the case without precise time-synchronization between the two nodes (the extension to the perfectly synchronous case is straightforward).
All proofs and derivations can be found in the appendix of our journal paper preprint \cite{DumphartTSP2021Preprint} but are omitted herein.

\subsubsection*{Paper Structure}
In \Cref{sec:SystemModel} we state the employed system model, geometric properties, and assumptions on synchronization and MPC extraction.
\Cref{sec:EstimateDist,sec:EstimateRelLoc} present the derived estimators for node distance and relative position, respectively.
The numerical performance evaluation is given in \Cref{sec:EvalSim}. \Cref{sec:Summary} concludes the paper.


	\section{System Model and Key Principles}
	\label{sec:SystemModel}
	The location of nodes A and B are written in Cartesian coordinates $\posA, \, \posB \in \bbR^3$ in an arbitrary reference frame. The location of node B relative to node A is characterized by
the relative position
$\d = \posB - \posA \in \bbR^3$
and the distance
$d  = \|\d\|$.
We consider the observers $o \in \{1, \ldots, \NObs\}$, $\NObs \geq 1$. The multipath channels between the observers and node A are characterized by the CIRs
$\cirA_o(\tau)$
and those between the observers and node B by
$\cirB_o(\tau)$.
All analytical statements require that the same set of MPCs is extracted from both $\cirA_o(\tau)$ and $\cirB_o(\tau)$.
This is easily fulfilled in practice for small $d$ and distinct MPCs. Large $d$, diffuse multipath, selective MPC occurrence and limited bandwidth will however pose challenges.
The number $K_o$ of extracted MPCs is left as unspecified design parameter.
The MPCs may or may not comprise the LOS path as well as reflected, scattered or diffracted paths \cite{WitrisalSPM2016}.

\begin{figure}[!ht]
\centering
\subfloat[multipath propagation from observer $o$ to A and B]{\centering
	\includegraphics[height=34mm]{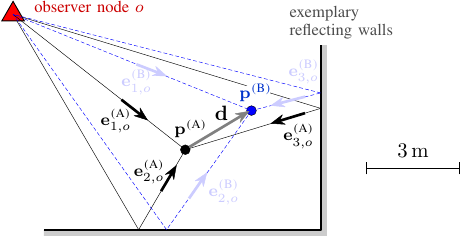}
	\label{fig:InterUserPaths}}
	\ \\[3.5mm]
\subfloat[CIRs between $o$ and A, between $o$ and B]{\centering
	\includegraphics[width=74mm]{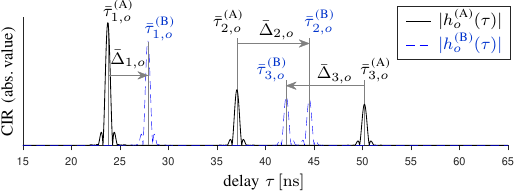}
	\label{fig:InterUserSignals}}
\caption{Considered wireless setup in an exemplary indoor environment with two walls, one observer ($\NObs = 1$), and $K_o = 3$ propagation paths. Here, $k=1$ is the LOS path and $k \in \{2,3\}$ are reflections. This CIR illustration assumes a raised-cosine pulse of $2\,\mathrm{GHz}$ bandwidth and no noise or interference.}
\label{fig:InterUserIntro}
\end{figure}

We use the MPC index $k \in \{ 1, 2, \ldots, K_o \}$ 
and the total MPC count $K = K_1 + \ldots + K_\NObs$.
The MPC delays are considered in terms of their true values
$\delayTrueA[k,o], \delayTrueB[k,o] \in \mathbb{R}_+$
and the observed values
$\delayMeasA[k,o], \delayMeasB[k,o] \in \mathbb{R}_+$.
Crucial quantities are the
\begin{align}
&\text{delay difference, true value:} &&
\delayDiffTrue[k,o] = \delayTrueB[k,o] - \delayTrueA[k,o] \, ,
\label{eq:DelayShiftDef_True}
\\
&\text{delay difference, measured:} &&
\delayDiff[k,o]     = \delayMeasB[k,o] - \delayMeasA[k,o]  \, .
\label{eq:DelayShiftDef_Meas}
\end{align}
The unit vectors
$\dirVectA[k,o], \dirVectB[k,o] \in \bbR^3$ denote the MPC directions. They describe the directions of arrival at $\posA\! , \posB$ if an observer is transmitting (cf. \Cref{fig:InterUserPaths}). Likewise, $-\dirVectA[k,o], -\dirVectB[k,o]$ are the directions of departure at $\posA\! , \posB$ if A, B are transmitting.


\vspace{-1.5mm}

We recall that a delay $\tau$ is caused by having traveled a path length $\cd\tau$ at the wave propagation velocity $c \approx 3 \cdot 10^8\unit{m/s}$ (the speed of light). This dependence gives rise to important \textit{geometric properties} regarding MPC dissimilarity between the two CIRs.
Each MPC $k,o$ fulfills the delay-difference bounds
\begin{align}
-d \leq \cd\delayDiffTrue[k,o] \leq d
\label{eq:DelayDiffBounds}
\end{align}
and, furthermore, two equalities on the relative position vector:
\begin{align}
\d &= \cd\delayTrueB[k,o] \dirVectB[k,o] - \cd\delayTrueA[k,o] \dirVectA[k,o]
\label{eq:VectorEquality}
\, , \\[1mm]
( \hspace{.1mm} \dirVectA[k,o] + \dirVectB[k,o] \hspace{.1mm} )\Tr \d
&= \cd\delayDiffTrue[k,o] \big(1 + (\dirVectA[k,o])\Tr \dirVectB[k,o] \,\big)
\, . 
\label{eq:ProjectionEquality}
\end{align}
A simple proof is given in \cite[Apdx.~A]{DumphartTSP2021Preprint}, based on the triangle between $\posA$, $\posB$, and the virtual source position of MPC $k,o$.
If
$d \ll \cd\delayTrueA[k,o]$,
$d \ll \cd\delayTrueB[k,o]$, and the MPC is not caused by a scatterer near A and B, then $\dirVectA[k,o] \approx \dirVectB[k,o]$ and in consequence
\begin{align}
(\dirVectA[k,o])\Tr \d \ \approx \
(\dirVectB[k,o])\Tr \d \ \approx \ \cd\delayDiffTrue[k,o]
\label{eq:ProjectionApproximation}
\end{align}
hold in good approximation. This is essentially a plane-wave assumption (PWA) in the vicinity of node A.

The properties \Cref{eq:DelayDiffBounds,eq:VectorEquality,eq:ProjectionEquality,eq:ProjectionApproximation} lay the foundation for the following estimators. A key strength is the formal absence of the observer positions and other environment specifics in the expressions.
The bounds \Cref{eq:DelayDiffBounds} show that the value range of $\{ \cd\delayDiff[k,o] \}$ is expressive of $d$, which will be utilized by the distance estimators in \Cref{sec:EstimateDist}.
Likewise, \Cref{eq:VectorEquality,eq:ProjectionEquality,eq:ProjectionApproximation} will be utilized by the relative position estimators in \Cref{sec:EstimateRelLoc}. We note that equation \Cref{eq:VectorEquality} readily provides an estimation rule for vector $\d$ from delay and direction of a single MPC, if accurate measurements thereof can actually be obtained.


For the measured MPC delays we consider the error model
$\delayMeasA[k,o] = \delayTrueA[k,o] + \errMeasA[k,o] + \errClockA[o]$
and
$\delayMeasB[k,o] = \delayTrueB[k,o] + \errMeasB[k,o] + \errClockB[o]$
where
$\errMeasA[k,o] , \errMeasB[k,o]$
are measurement errors due to noise, interference, limited bandwidth, clock jitter, receiver resolution, and other imperfections \cite{DardariPIEEE2009}.
The clock offsets
$\errClockA[o]$ between A and $o$ and 
$\errClockB[o]$ between B and $o$ occur because we do not assume precise time synchronization, neither between A and B nor between the observers. 
However, we assume that the setup is able to conduct the necessary channel estimation steps in quick succession, such that the clock drift is negligible over the duration of the entire process, i.e. while recording all the received signals.
A consequence is the property
$\errClockB[o] - \errClockA[o] = \errClock$
for the clock offset $\errClock$ between A and B, which specifically does not depend on the observer index $o$ (without this property, the presented estimators apply with straightforward adaptations).
This property yields a particularly simple \textit{signal model} for the measured delay differences
$\delayDiff[k,o]$ 
from \Cref{eq:DelayShiftDef_Meas},
\begin{align}
\delayDiff[k,o] 
&= \delayDiffTrue[k,o] + \errMeas[k,o] + \errClock \, .
\label{eq:SignalModel}
\end{align}
The measurement error is given by
$\errMeas[k,o] = \errMeas[k,o]\AnnotateNode{B} - \errMeas[k,o]\AnnotateNode{A} \in \bbR$
and is considered as random variable. We note that potential biases are compensated by the difference. Furthermore, because of the many different influences in
$\errMeas[k,o]\AnnotateNode{B}$ and $\errMeas[k,o]\AnnotateNode{A}$, a zero-mean Gaussian distribution would be a reasonable assumption for $\errMeas[k,o]$ (central limit theorem).
The clock offset $\errClock \in \bbR$ is considered unknown but non-random.


Finally, we comment on the \textit{MPC association problem} between $\cirA_o(\tau)$ and $\cirB_o(\tau)$. This is a delicate signal processing problem when the path delays and amplitudes are the only available MPC-identifying features.
The problem is apparent in \Cref{fig:InterUserSignals} where $\delayTrueA[2,o] < \delayTrueA[3,o]$ while $\delayTrueB[3,o] < \delayTrueB[2,o]$. This shows that sorting the delays in ascending order may not yield the MPC association. This is to be expected unless $d/c$ is much smaller than the delay spread. On the other hand, MPC association can be trivial if $d$ is small.
This paper contains estimators for the cases of known and unknown MPC association.

	\section{Distance Estimators}
	\label{sec:EstimateDist}
	This section presents estimators of the inter-node distance $d$ from measured delay differences $\delayDiff[k,o] = \delayDiffTrue[k,o] + \errMeas[k,o] + \errClock$ as defined in \Cref{eq:SignalModel}. The estimators do not use or require the MPC directions.
The estimation-theoretic properties (MLE and MVUE) relate to the following employed assumptions regarding rich multipath propagation:
(i) each MPC direction $\dirVectA[k,o]$ is random and has uniform distribution on the 3D unit sphere,
(ii) the random directions $\dirVectA[k,o]$ are statistically independent for different MPCs $k,o$, and
(iii) the PWA \Cref{eq:ProjectionApproximation} is assumed to be exact.
These assumptions actually result in uniform distributions
$\cd\delayDiffTrue[k,o] \iid \ \mathcal{U}(-d,d)$ for the delay differences \cite{DumphartICC2019}.

%
We assume measurement errors $\errMeas[k,o]$ with known distribution and statistical independence between different MPCs $k,o$.

\Cref{sec:EstimateDistWithAssoc} assumes that the MPC association was established correctly by a preceding signal processing step (these estimators are restatements of the results in \cite{DumphartICC2019}, included for coherence). In contrary, \Cref{sec:EstimateDistNoAssoc} operates without any such knowledge of the MPC association.


	\subsection{With Known MPC Association}
	\label{sec:EstimateDistWithAssoc}
	Given are the delay differences $\delayDiff[k,o]$ subject to measurement errors $\errMeas[k,o]$ and an unknown clock offset $\errClock$. The joint maximum-likelihood estimate (MLE) of distance and clock offset is given by the maximization problem \cite[Apdx.~B-1]{DumphartTSP2021Preprint}
\begin{align}
& ( \hat d \MLE\AssAsyn , \hat \errClock \MLE\AssAsyn )
\in \argmax_{\Hypo{d} \in \bbR_+, \Hypo{\errClock} \in \bbR} \,\f{1}{\Hypo{d}^K \!} \prod_{o=1}^{\NObs} \prod_{k=1}^{K_o}
I_{k o}\big( \delayDiff[k,o] \!-\! \Hypo{\errClock}, \Hypo{d} \, \big)
, \label{eq:distMLE_general} \\
& I_{k o}\big(\, \bullet \, ,\Hypo{d} \,\big) =
F_{\errMeas[k,o]}\big(\bullet + \tfrac{\Hypo{d}}{c} \,\big) -
F_{\errMeas[k,o]}\big(\bullet - \tfrac{\Hypo{d}}{c} \,\big) .
\label{eq:SoftIndicFunc}
\end{align}
The free variables $\Hypo{d}$ and $\Hypo{\errClock}$ represent distance and clock-offset hypothesis, respectively, and $F_{\errMeas[k,o]}$ is the cumulative distribution function (CDF) of the measurement error $\errMeas[k,o]$.
The non-random $\errClock$ is necessarily included as nuisance parameter.

The term $I_{k o}$ can be regarded as soft indicator function that evaluates the set membership $\cd (\delayDiff[k,o] - \errClock) \in [-\Hypo{d},\Hypo{d}\,]$.
For the case of Gaussian errors $\errMeas[k,o] \sim \mathcal{N}(0 , \sigma_{k o}^2)$, the CDF is described by the \mbox{$Q$-function}, $F_{\errMeas[k,o]}(x) = 1-Q(x/\sigma_{k o})$.
Examples of the two-dimensional likelihood function, i.e. of the maximization objective function in \Cref{eq:distMLE_general}, are given in \Cref{fig:LHF_Asyn,fig:LHF_Asyn_Gauss}. They concern the setup in \Cref{fig:InterUserPaths}.

\newcommand{\myFigHeight}{27mm}
\begin{figure}[!ht]\centering
    \subfloat[known assoc., no meas. error]{
        \includegraphics[height=\myFigHeight]{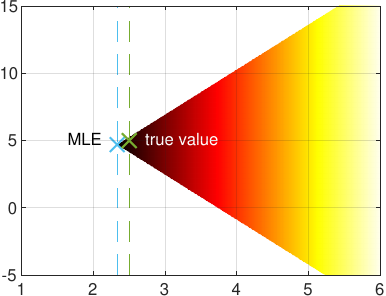}
        \ \ \ \,
        \put(-122,10){\rotatebox{90}{\tiny{clock offset hypothesis $\Hypo{\errClock}$ [ns]}}}
        \put(-89,-8){\tiny{distance hypothesis $\Hypo{d}$ [m]}}
        \label{fig:LHF_Asyn}}
    \subfloat[known assoc., $\sigma = 1\unit{ns}$]{
        \includegraphics[height=\myFigHeight]{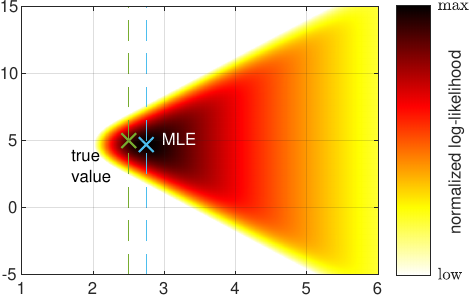}
        \put(-128,10){\rotatebox{90}{\tiny{clock offset hypothesis $\Hypo{\errClock}$ [ns]}}}\put(-95,-8){\tiny{distance hypothesis $\Hypo{d}$ [m]}}
        \label{fig:LHF_Asyn_Gauss}}
    \\
    \subfloat[no MPC assoc., no meas. error]{
        \includegraphics[height=\myFigHeight]{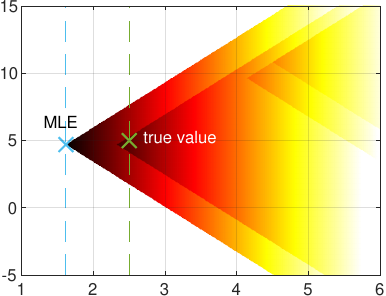}
        \ \ \ \,
        \put(-122,10){\rotatebox{90}{\tiny{clock offset hypothesis $\Hypo{\errClock}$ [ns]}}}
        \put(-89,-8){\tiny{distance hypothesis $\Hypo{d}$ [m]}}
        \label{fig:LHF_Asyn_NoAssoc}}
    \subfloat[no MPC assoc., $\sigma = 1\unit{ns}$]{
        \includegraphics[height=\myFigHeight]{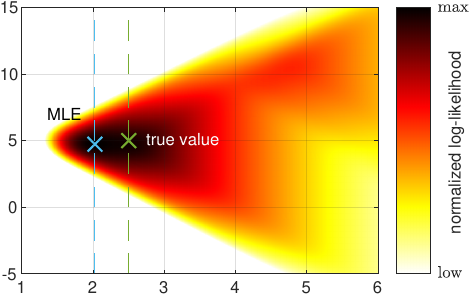}
        \put(-128,10){\rotatebox{90}{\tiny{clock offset hypothesis $\Hypo{\errClock}$ [ns]}}}
        \put(-95,-8){\tiny{distance hypothesis $\Hypo{d}$ [m]}}
        \label{fig:LHF_Asyn_NoAssoc_Gauss}}
\caption{Likelihood function of distance (abscissa) and clock offset (ordinate) given observed delay differences $\{ \delayDiff[k,o] \}$, cf. \Cref{eq:distMLE_general}.
Shown are cases with and without MPC association and measurement errors (Gaussian distribution). The evaluation is done for the setup in \Cref{fig:InterUserIntro} with the same CIRs with $K = 3$ and true value $d = 2.5\unit{m}$ (additionally $\errClock = 5\unit{ns}$ is assumed).}
\label{fig:LHF_Asyn_AllCases}
\end{figure}
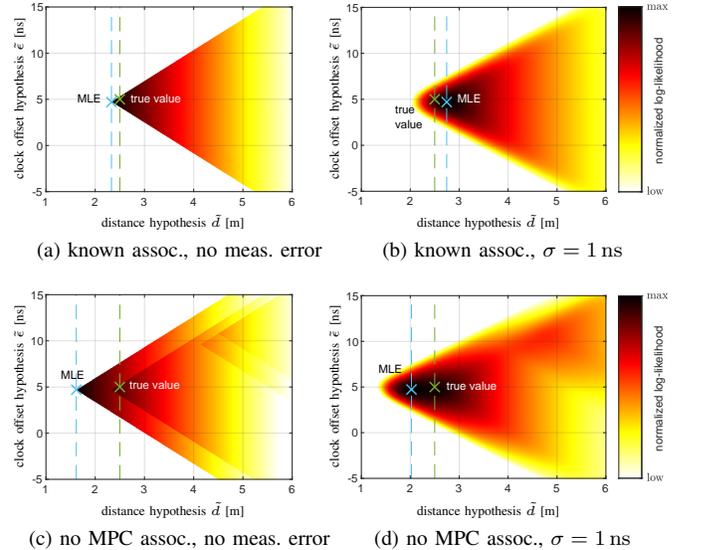

The optimization problem \Cref{eq:distMLE_general} can be tackled with numerical methods such as iterative gradient-based solvers (a general closed-form solution is unavailable). The likelihood function is non-concave in general and also in the Gaussian case.

Consider the special case of zero measurement errors ($\errMeas[k,o] \equiv 0$). Here
the actual indicator function $I_{k o}(\bullet,d) = \IndFunc_{[-d/c,d/c]}(\bullet)$ applies, the likelihood function attains a distinct structure (see \Cref{fig:LHF_Asyn}), and the MLE problem \Cref{eq:distMLE_general} now has the closed-form solution $\hat d \MLE
= \f{c}{2} ( \max_{k,o} \delayDiff[k,o] - \min_{k,o} \delayDiff[k,o] )$, as shown in
\cite[Apdx.~B-2]{DumphartTSP2021Preprint}, \cite{DumphartICC2019}.
This estimator has a negative bias; it underestimates with probability $1$.
A bias-correction actually yields the MVUE, \cite{DumphartICC2019}
\begin{align}
&\hat d \MVUE = \f{K+1}{K-1}\ \f{c}{2} \Big( \max_{k,o} \delayDiff[k,o] - \min_{k,o} \delayDiff[k,o] \Big)
\, , \label{eq:distMVUE} \\
& \hat \errClock\MVUE = \hat \errClock\MLE\AssAsyn = \f{1}{2} \Big( \max_{k,o} \delayDiff[k,o] + \min_{k,o} \delayDiff[k,o] \Big)
\, . \label{eq:EpsMVUE} 
\end{align}
If $\errClock = 0$ due to precise a-priori time synchronization, then
$\hat d \MLE = c \cdot \max_{k,o} |\delayDiff[k,o]|$
and
$\hat d \MVUE = \f{K+1}{K} \hat d \MLE$. \cite{DumphartICC2019}


	\subsection{With Unknown MPC Association}
	\label{sec:EstimateDistNoAssoc}
    We now assume that, for any given $o$, the association between the delays
$\{ \delayMeasA[1,o] \, , \ldots \, , \, \delayMeasA[K_o,o] \}$
and
$\{ \delayMeasB[1,o] \, , \ldots \, , \, \delayMeasA[K_o,o] \}$
is unavailable.
A distance estimator now faces the problem that (without any prior knowledge) any MPC association is eligible. Hence, any conceivable delay-difference $\delayMeasB[\perm(k),o] - \delayMeasA[k,o]$ with any choice of permutation $k' = \perm(k)$ is eligible. We refer to a permutation $\perm \in \permSet[K_o]$ as a bijective map from and to $\{1,\ldots,K_o\}$; it serves as formal representation of MPC association.
A sorting permutation which establishes
$\delayMeasA[      k ,o] < \delayMeasA[      l ,o]$ iff
$\delayMeasB[\perm(k),o] < \delayMeasB[\perm(l),o]$
will likely be the correct association if $d/c$ is much smaller than the channel delay spread. This criterion however cannot be evaluated without prior knowledge on $d$.

For unknown MPC association, we find that the joint MLE of distance and clock offset is given by
\begin{multline}
\left( \hat d \MLE\NoAssAsyn, \hat\errClock \MLE\NoAssAsyn \right) 
\in
\argmax_{\Hypo{d} \in \R_+, \, \Hypo{\errClock} \in \R}
\\
\frac{1}{\Hypo{d}^K} \prod_{o=1}^\NObs \sum_{\perm \in \permSet[K_o]} \prod_{k=1}^{K_o}
I_{ko}\big( \delayMeasB[\perm(k),o] - \delayMeasA[k,o] - \Hypo{\errClock}, \Hypo{d} \,\big)
\label{eq:distMLENoAss_}
\end{multline}
with the soft indicator function $I_{ko}$ from \Cref{eq:SoftIndicFunc}. The derivation can be found in \cite[Apdx.~C]{DumphartTSP2021Preprint}.
Examples of the likelihood function are given in \Cref{fig:LHF_Asyn_NoAssoc,fig:LHF_Asyn_NoAssoc_Gauss}.
The estimates can be obtained by (attempting to) compute the global solution of the optimization problem \Cref{eq:distMLENoAss_} with a numerical solver, e.g., an iterative gradient-based algorithm with a multistart approach.

If $\errMeas[k,o] \equiv 0$ then the likelihood function attains a distinct structure. Then it suffices to evaluate the likelihood at a finite set of MLE candidate values
given by the peaks of wedges and intersections of wedge-borders, as seen in \Cref{fig:LHF_Asyn_NoAssoc}.

    \section{Position Estimators}
	\label{sec:EstimateRelLoc}
	This section presents estimators of the relative position vector $\d = \posB - \posA$ from measured delay differences $\delayDiff[k,o]$ or, alternatively, directly from measured MPC delays $\delayMeasA[k,o]$ and $\delayMeasB[k,o]$. These estimators do use the MPC directions $\dirVectA[k,o], \dirVectB[k,o]$ and thus require their availability, e.g., by measuring them during channel estimation with the use of antenna arrays at A and B. The estimators assume that the MPC association was established correctly by a preceding signal processing step, e.g. by the scheme presented later in \Cref{sec:EstimateRelLocNoAssoc}.

	
	\subsection{From Delay Differences}
	\label{sec:EstimateRelLocFromDelta}
	We define the stacked vector and matrix quantities
\begin{align}
\delayDiff
&= [ \, \delayDiff[1,1] \ldots \delayDiff[K_1,1] \ \delayDiff[1,2] \ldots \delayDiff[K_\NObs,\NObs] ]\Tr
\! \! \! 
&& \in \bbR^{K \times 1}
, \label{eq:DelayShiftStackVector} \\
\errMeas
&= [ \ \, \errMeas[1,1] \ldots \ \errMeas[K_1,1] \ \ \errMeas[1,2] \ldots \, \,\errMeas[K_\NObs,\NObs] ]\Tr \!\!
&& \in \bbR^{K \times 1}
, \label{eq:ErrorStackVector} \\
\E &= \hspace{-.5mm}
\mtx{cc}{
\! {\bf s}_{1,1} \,\ldots\, {\bf s}_{K_1,1} \!\! & \!\! {\bf s}_{1,2} \,\ldots\, {\bf s}_{K_\NObs, \NObs} \! \! \\
\!\!\! 1 \,\ \ldots \, \ \ 1 \!\! & \!\!\!\!\!\!\!\!\!\! 1 \ \ \ \ldots \ \ 1
} &&\in \bbR^{4 \times K}
, \label{eq:EMatrix} \\
{\bf s}_{k,o} &= 
\f{1}{1 + (\dirVectA[k,o])\Tr \dirVectB[k,o]}
\left(\, \dirVectA[k,o] + \dirVectB[k,o] \,\right)
&& \in \bbR^{3 \times 1}
\label{eq:sVectorForEMatrix}
\end{align}
which establish the simple relation $\delayDiff = \f{1}{c} \E\Tr[\d\Tr , \cd\errClock]\Tr + \errMeas$.
Moreover, \Cref{eq:sVectorForEMatrix} has the property ${\bf s}_{k,o}\Tr \d = \cd\delayDiffTrue[k,o]$. This is a restatement of the projection property \Cref{eq:ProjectionEquality} which forms the basis of the following scheme.

As observations we consider $\delayDiff[k,o]$ as well as the MPC directions $\dirVectA[k,o]$ and $\dirVectB[k,o]$.
The relative position MLE is given by the unconstrained four-dimensional optimization problem
$\argmax
f_{\errMeas} ( \delayDiff - \f{1}{c} \E\Tr[\Hypo{\d}\Tr , \cd\Hypo{\errClock}]\Tr )$
over
$\Hypo{\d} \in \bbR^3$ and $\Hypo{\errClock} \in \bbR$.
It uses the joint PDF $f_{\errMeas}$ of measurement errors,
cf. \cite[Apdx.~D-1]{DumphartTSP2021Preprint}.

If $f_{\errMeas}$ is unavailable, then still the least-squares estimate (LSE) can be computed. It is given by the simple formula
\begin{align}
\mtx{c}{\hat\d\LSE\AssViaDiff \\[1mm] \!
c \cdot \hat\errClock\LSE\AssViaDiff \!}
&= 
\left(\E\E\Tr \right)^{-1} \E\, (\cd\delayDiff)
\, . \label{eq:displLSEAssViaDiff}
\end{align}
For the special case $\errMeas \sim \calN({\bf 0}, \sigma^2 \,\eye_K)$, the LSE \Cref{eq:displLSEAssViaDiff} is also the MLE and the MVUE. For a general Gaussian-distributed $\errMeas \sim \calN(\boldsymbol\mu, \boldsymbol\Sigma)$, the MLE and MVUE is given by $( \E\,\boldsymbol\Sigma^{-1} \E\Tr )^{-1} \E \, \boldsymbol\Sigma^{-1} (\cd\delayDiff - c\boldsymbol\mu)$. \cite[Thm~4.2 and Thm~7.5]{Kay1993}.

When only the directions $\dirVectA[k,o]$ but not $\dirVectB[k,o]$ are available, then the PWA
$\dirVectA[k,o] \approx \dirVectB[k,o] \ \Rightarrow\ {\bf s}_{k,o} \approx \dirVectA[k,o]$
allows to still use the MLE or LSE with hardly any accuracy loss at small $d$.
To that effect, we define
\begin{align}
\hat\d\LSE\AssViaDiffPWA := \hat\d\LSE\AssViaDiff \Big|_{{\bf s}_{k,o} = \,\dirVectA[k,o]} \, .
\label{eq:displLSEAssViaDiffPWA}
\end{align}


	\subsection{Directly From Delays}
	\label{sec:EstimateRelLocFromTau}
	Based on the property $\d = \cd\delayTrueB[k,o] \dirVectB[k,o] - \cd\delayTrueA[k,o] \dirVectA[k,o]$ from \Cref{eq:VectorEquality},
we study an alternative scheme which uses the delays directly instead of their difference.
In particular, we consider estimation of vector $\d$ from measured delays
$\delayMeasA[k,o]$ and $\delayMeasB[k,o]$ that are subject to and measurement errors as well as
clock offsets $\errClock_o\AnnotateNodeLOWER{A}$ and $\errClock_o\AnnotateNodeLOWER{B} = \errClock_o\AnnotateNodeLOWER{A} - \errClock$, respectively.
We find that the joint LSE of $\d$ and all relevant clock offsets is given by
\begin{align}
& \left[
(\hat\d\LSE\AssViaTau)\Tr \! ,
\cd\hat\errClock ,
\cd\errClockAEstimate[1] ,
\ldots ,
\cd\errClockAEstimate[\NObs]
\right]\Tr
\! = 
( {\bf G}\Tr {\bf G} )^{-1} {\bf G}\Tr {\bf t}
\label{eq:EstimateRelLocViaTau}
\end{align}
whereby ${\bf G} \in \bbR^{(3K) \times (4 + \NObs)}$ and ${\bf t} \in \bbR^{(3K) \times 1}$ are defined as
\begin{align}
{\bf G} = &
\mtx{llccc}{
\eye_3 & \dirVectB[1,1]   & \dirVectB[ 1 ,1]\!-\dirVectA[ 1 ,1] \\[-2mm]
\vdots & \,\vdots         &    \vdots                           & & {\bf 0} \\
\eye_3 & \dirVectB[K_1,1] & \dirVectB[K_1,1]\!-\dirVectA[K_1,1] \\[-2mm]
\vdots & \,\vdots & & \!\!\!\!\ddots\!\!\!\! \\
\eye_3 & \dirVectB[1,\NObs] & & & \dirVectB[ 1 ,\NObs]\!-\dirVectA[ 1 ,\NObs] \\[-2mm]
\vdots & \,\vdots       & {\bf 0} & & \vdots \\
\eye_3 & \dirVectB[K_\NObs,\NObs] & & & \!\!\!\dirVectB[K_\NObs,\NObs]\!-\dirVectA[K_\NObs,\NObs]\!\!\! } 
, \nonumber \\
{\bf t} = &
\mtx{c}{
\cd\delayMeasB[1,1] \dirVectB[1,1] - \cd\delayMeasA[1,1] \dirVectA[1,1] \\[-1.5mm]
\vdots \\[-1mm]
\!\!
\cd\delayMeasB[K_\NObs,\NObs] \dirVectB[K_\NObs,\NObs] - \cd\delayMeasA[K_\NObs,\NObs] \dirVectA[K_\NObs,\NObs]
\!\!}
. \label{eq:EstimateRelLocViaTauDetails}
\end{align}
The derivation can be found in \cite[Apdx.~D-2]{DumphartTSP2021Preprint}.
The approach relies on accurately measured MPC directions $\dirVectA[k,o]$, $\dirVectB[k,o]$ and is fundamentally incompatible with the PWA \Cref{eq:ProjectionApproximation}.

If $\errClockA[o] = 0$ and $\errClockB[o] = 0$ are a-priori established for all $o$, then
$\hat\d\LSE\AssViaTauSync =$
$
\f{1}{K}
\sum_{o=1}^\NObs \sum_{k=1}^{K_o}
\cd\delayMeasB[k,o] \dirVectB[k,o] - 
\cd\delayMeasA[k,o] \dirVectA[k,o]
$ applies. This is just the componentwise mean of \Cref{eq:VectorEquality} over all MPCs.
	
	\subsection{Establishing the MPC Association}
	\label{sec:EstimateRelLocNoAssoc}
	The relative position estimators stated in \Cref{sec:EstimateRelLocFromDelta,sec:EstimateRelLocFromTau}
assume the availability of the MPC association and of the MPC directions $\dirVectA[k,o]$ and $\dirVectB[k,o]$. The MPC directions are particularly useful for reconstructing the MPC association in case it is a-priori unknown, which is the topic of this subsection.

We assume that the MPC directions $\dirVectA[k,o]$ and $\dirVectB[k,o]$ are stated within the same frame of reference (this could be enforced by solving an orthogonal Procrustes problem).
The MPC association relating to observer $o$ is formalized in terms of a permutation $k' = \pi_o(k)$ with $k,k' \in \{1,\ldots,K_o\}$.
As a tool to reconstruct the MPC association given the MPC directions and delays, we propose the geometry-inspired data-fitting rule
\begin{align}
\hat\perm_o &= \argmin_{\perm \in \permSet[K_o]} \sum_{k=1}^{K_o} J_o(k,\perm(k))
\label{eq:AssocEstim} \, , \\
J_o(k,l) &=
\big\| {\dirVectB[{l,o}]} \! - \dirVectA[k,o] \big\|^2
+ \lambda^2
\big| \delayMeasB[l,o]\! - \mu_o\AnnotateNode{B}\! - \delayMeasA[k,o]\! - \mu_o\AnnotateNode{A} \big|^2
\nonumber
\end{align}
where $\mu_o\AnnotateNode{\textbullet} = \f{1}{K_o} \sum_{k=1}^{K_o} \tau_{k,o}\AnnotateNode{\textbullet}$ is the mean delay.
The regularization constant $\lambda^2$ balances cost contributions by directions and delays.
With large $\lambda$, \Cref{eq:AssocEstim} tends to associate the MPCs by sorting the delays in ascending order.
A sensible choice is given by $\lambda = 1/\tauRMS$ where $\tauRMS$ is the channel delay spread.
A reconstructed association $\hat\perm_o$ between MPCs $\bullet_{k,o}\AnnotateNode{A}$ and $\bullet_{\hat\perm_o{(k)},o}\AnnotateNode{B}$ allows to use any estimator from \Cref{sec:EstimateDistWithAssoc,sec:EstimateRelLocFromDelta,sec:EstimateRelLocFromTau}.

The cost function $J_o(k,l)$ bears similarities to the well-known optimal subpattern assignment (OSPA) metric \cite{LeitingerTWC2019}.
The optimization problem \Cref{eq:AssocEstim} is a linear assignment problem and can thus be solved efficiently with the Hungarian method.
The framework of linear assignment problems can handle MPCs $\bullet_{k,o}\AnnotateNodeLOWER{A}$ without a corresponding $\bullet_{k,o}\AnnotateNodeLOWER{B}$. It is furthermore able to detect and reject bad associations: large angles between $\dirVectA[k,o]$ and $\dirVectB[lo]$ indicate an incorrect association. We implement this notion by setting $J_o(k,l) = \infty$ if the angle exceeds $30^\circ$.

	\section{Numerical Performance Evaluation}
	\label{sec:EvalSim}
	\newcommand\SimPlotsInnerWidth{.5\columnwidth}  
\newcommand\SimPlotsInnerHeight{.4\columnwidth} 

\newcommand\SimPlotsLineWidth{0.5pt}
\newcommand\SimPlotsYMaxDist{1.2}
\newcommand\SimPlotsYMaxPos{1.2}

\newcommand\SimPlotsGrid{false}
\newcommand\SimPlotsGridLineWidth{0.5pt}
\newcommand\SimPlotsGridColor{gray!18}

\newcommand\SimPlotsLegendLocationDist{(0.06,1)}
\newcommand\SimPlotsLegendLocationPos{(0.07,1)}

\newcommand\SimPlotsLegendCaseA{MV} 
\newcommand\SimPlotsLegendCaseC{NA} 
\newcommand\SimPlotsLegendCaseD{SO} 
\newcommand\SimPlotsLegendCaseE{DD} 
\newcommand\SimPlotsLegendCaseF{PWA} 
\newcommand\SimPlotsLegendCaseG{DDN} 
\newcommand\SimPlotsLegendCaseH{TAU} 
\newcommand\SimPlotsLegendCaseI{TNA} 
\newcommand\SimPlotsTextCaseA{$\hat d\MVUE$ from \Cref{eq:distMVUE}} 
\newcommand\SimPlotsTextCaseC{$\hat d\MLE\NoAssAsyn$ from \Cref{eq:distMLENoAss_} (unknown MPC assoc.)}
\newcommand\SimPlotsTextCaseD{$\hat d\MVUE$ from \Cref{eq:distMVUE}, assoc. estimated via $\tau$-sorting}
\newcommand\SimPlotsTextCaseE{$\hat\d\LSE\AssViaDiff$ from \Cref{eq:displLSEAssViaDiff}}
\newcommand\SimPlotsTextCaseF{$\hat\d\LSE\AssViaDiffPWA$ from \Cref{eq:displLSEAssViaDiffPWA} (plane-wave assumption)}
\newcommand\SimPlotsTextCaseG{$\hat\d\LSE\AssViaDiff$ from \Cref{eq:displLSEAssViaDiff}, assoc. estimated via \Cref{sec:EstimateRelLocNoAssoc}}
\newcommand\SimPlotsTextCaseH{$\hat\d\LSE\AssViaTau$ from \Cref{eq:EstimateRelLocViaTau}}
\newcommand\SimPlotsTextCaseI{$\hat\d\LSE\AssViaTau$ from \Cref{eq:EstimateRelLocViaTau}, assoc. estimated via \Cref{sec:EstimateRelLocNoAssoc}}

\newcommand\SimPlotsALineStyle{solid}
\newcommand\SimPlotsAMarkSize{3pt}
\newcommand\SimPlotsAMarker{triangle}
\newcommand\SimPlotsARotate{180}

\newcommand\SimPlotsCLineStyle{dotted}
\newcommand\SimPlotsCMarkSize{\SimPlotsAMarkSize}
\newcommand\SimPlotsCMarker{triangle*}

\newcommand\SimPlotsDLineStyle{dashed}
\newcommand\SimPlotsDMarkSize{3.0pt}
\newcommand\SimPlotsDMarker{asterisk}

\newcommand\SimPlotsELineStyle{solid}
\newcommand\SimPlotsEMarkSize{2.5pt}
\newcommand\SimPlotsEMarker{square}

\newcommand\SimPlotsFLineStyle{dashdotted}
\newcommand\SimPlotsFMarkSize{1.0pt}
\newcommand\SimPlotsFMarker{*}

\newcommand\SimPlotsGLineStyle{dotted}
\newcommand\SimPlotsGMarkSize{3.5pt}
\newcommand\SimPlotsGMarker{+}

\newcommand\SimPlotsHLineStyle{solid}
\newcommand\SimPlotsHMarkSize{3.5pt}
\newcommand\SimPlotsHMarker{diamond}

\newcommand\SimPlotsILineStyle{dashed}
\newcommand\SimPlotsIMarkSize{3.5pt}
\newcommand\SimPlotsIMarker{x}

We evaluate the estimators' accuracy numerically via random sampling of MPC parameters. The statistical assumptions described in the following are characteristic of dense indoor multipath channels.
Any observer is at a distance of $5\unit{m}$ from $\posA$, giving a minimum delay of $\delayTrueA[k o] \geq \tau_\mathrm{min} = 16.7\unit{ns}$.
The excess delay is sampled according to the statistical channel model for indoor multipath propagation by Saleh and Valenzuela \cite{SalehJSAC1987} with
$20\unit{ns}$ cluster mean,
$10\unit{ns}$ delay mean,
$\f{1}{60}$ cluster loss, and
$\f{1}{20}$ delay loss.
This results in a double-exponential power-delay profile with
an RMS delay spread of $\sigma_\tau = 26.3\unit{ns}$,
a mean excess delay of $\bar\tau = 40.5\unit{ns}$, and
a mean delay of $\EVSymb[\delayTrueA[k o]] = \tau_\mathrm{min} + \bar\tau = 57.2\unit{ns}$.
We expect problems with MPC association unless $d \ll \cd\sigma_\tau = 7.9\unit{m}$ and, likewise, we expect the PWA \Cref{eq:ProjectionApproximation} to be accurate for $d \ll c\cdot\EVSymb[\delayTrueA[k o]] = 17.2\unit{m}$.
Additionally, the MPC directions $\dirVectA[k o]$ are independently drawn from a uniform distribution over the 3D unit sphere.
We set $\posB = \posA + \d$ with $\d = [d,0,0]\Tr$. Then $\dirVectB[k o]$ is obtained by normalizing $\d + \cd\delayTrueA[k o]\dirVectA[k o]$.
Delay measurement errors are sampled according to
$\errMeasA[k o],\errMeasB[k o] \iid \calN(0,\f{\sigma^2}{2})$,
giving $\errMeas[k o] \iid \calN(0,\sigma^2)$.
Any MPC direction ${\dirVect}_{k o}^{(\,.\,)}$ used by a relative position estimator (i.e. a measured direction) is assumed to deviate from its true value by an angle $\alpha \sim \calN(0,\sigma_\text{dir}^2)$; the unit vector ${\dirVect}_{k o}^{(\,.\,)}$ is then uniformly sampled from the circle defined by the value of $\alpha$.

We assume the values $d = 2\unit{m}$, $\sigma = 0.2\unit{ns}$, $\sigma_\text{dir}= 0$ unless either variable defines the abscissa of a graph. A legend of the evaluated estimators is given in \Cref{tab:SimPlot}.

The experiment in \Cref{fig:RandomSamplingRMSE_d2m} studies the effect of varying $d$ and $\sigma_\text{dir}$. It considers $K_o = 4$ MPCs for each of three observers $o \in \{1,2,3\}$, giving a total of $K = 12$ MPCs.

\Cref{fig:RandomSamplingRMSE_vs_d_Dist} shows that distance estimation suffers a significant accuracy loss when the MPC association is unknown (cases \SimPlotsLegendCaseC{} and \SimPlotsLegendCaseD{}).
The simple \SimPlotsLegendCaseD{} scheme, which just applies $\hat d\MVUE$ after associating the MPCs by sorting the delays in ascending order, surprisingly outperforms the sophisticated estimator $\hat d\MLE\NoAssAsyn$ (\SimPlotsLegendCaseC{}). One reason for that is the lack of bias-correction in $\hat d\MLE\NoAssAsyn$, which could be addressed by future work.

Regarding the position estimators, \Cref{fig:RandomSamplingRMSE_vs_d_Pos} demonstrates great accuracy for $\d\LSE\AssViaDiff$ (case \SimPlotsLegendCaseE{}) and even more so for $\d\LSE\AssViaTau$ (case \SimPlotsLegendCaseH{}). In this experiment they are limited only by the small assumed delay error $\sigma$. Even with the PWA or with reconstructed MPC association (scheme from \Cref{sec:EstimateRelLocNoAssoc}, cases \SimPlotsLegendCaseF{} and \SimPlotsLegendCaseI{}) the accuracy is great up to distances close to the observer distance (here $5\unit{m}$).

\newcommand\LegendSymb[4]{\begin{tikzpicture}
\begin{axis}[%
width=23mm, height=20mm, at={(0,0)},
axis line style={draw=none}, xtick=\empty, ytick=\empty,
xmin=0, xmax=2, ymin=-1, ymax=1,
axis background/.style={fill=white},
axis x line*=bottom, axis y line*=left,
]
\addplot [color=black, #1, line width=\SimPlotsLineWidth]
  table[row sep=crcr]{0 0\\2 0\\};
\addplot [color=black, line width=\SimPlotsLineWidth, mark size=#2, mark=#3, mark options={solid, black,rotate=#4}]
  table[row sep=crcr]{1 0\\};
\end{axis}
\end{tikzpicture}}
\begin{table}
\vspace{0.04in}
\centering
\newcommand\myDist{-1.0mm}
\newcolumntype{L}[1]{>{\arraybackslash}m{#1}}
\begin{tabular}{cc|L{.65\columnwidth}}
\hline
\ \\[-2.7mm]
\multicolumn{3}{c}{distance estimators} \\[.5mm]\hline
& \\[-2.7mm]
\LegendSymb{\SimPlotsALineStyle}{\SimPlotsAMarkSize}{\SimPlotsAMarker}{\SimPlotsARotate}
& \SimPlotsLegendCaseA & \SimPlotsTextCaseA \\[\myDist]
\LegendSymb{\SimPlotsCLineStyle}{\SimPlotsCMarkSize}{\SimPlotsCMarker}{0}
& \SimPlotsLegendCaseC & \SimPlotsTextCaseC \\[\myDist]
\LegendSymb{\SimPlotsDLineStyle}{\SimPlotsDMarkSize}{\SimPlotsDMarker}{0}
& \SimPlotsLegendCaseD & \SimPlotsTextCaseD \\\hline
\ \\[-2.7mm]
\multicolumn{3}{c}{position estimators} \\[.5mm]\hline
& \\[-2.7mm]
\LegendSymb{\SimPlotsELineStyle}{\SimPlotsEMarkSize}{\SimPlotsEMarker}{0}
& \SimPlotsLegendCaseE & \SimPlotsTextCaseE \\[\myDist]
\LegendSymb{\SimPlotsGLineStyle}{\SimPlotsGMarkSize}{\SimPlotsGMarker}{0}
& \SimPlotsLegendCaseG & \SimPlotsTextCaseG \\[\myDist]
\LegendSymb{\SimPlotsFLineStyle}{\SimPlotsFMarkSize}{\SimPlotsFMarker}{0}
& \SimPlotsLegendCaseF & \SimPlotsTextCaseF \\[\myDist]
\LegendSymb{\SimPlotsHLineStyle}{\SimPlotsHMarkSize}{\SimPlotsHMarker}{0}
& \SimPlotsLegendCaseH & \SimPlotsTextCaseH \\[\myDist]
\LegendSymb{\SimPlotsILineStyle}{\SimPlotsIMarkSize}{\SimPlotsIMarker}{0}
& \SimPlotsLegendCaseI & \SimPlotsTextCaseI
\end{tabular}
\vspace{1mm}
\caption{Legend of evaluated estimators.}
\label{tab:SimPlot}
\end{table}

\begin{figure}
\centering
\subfloat[]{\centering\label{fig:RandomSamplingRMSE_vs_d_Dist}
\resizebox{.48\columnwidth}{!}{
%
%
%
\begin{tikzpicture}

\begin{axis}[%
width=\SimPlotsInnerWidth,
height=\SimPlotsInnerHeight,
at={(0,0)},
scale only axis,
xmin=0,
xmax=8,
xtick={0, 1, 2, 3, 4, 5, 6, 7, 8},
xlabel style={font=\color{white!15!black}},
xlabel={inter-node distance $d$ [m]},
ymin=0,
ymax=\SimPlotsYMaxDist,
ytick distance=.2,
ylabel style={font=\color{white!15!black}},
ylabel={distance RMSE [m]},
axis background/.style={fill=white},
axis x line*=bottom,
axis y line*=left,
xmajorgrids=\SimPlotsGrid,
ymajorgrids=\SimPlotsGrid,
grid style={line width=\SimPlotsGridLineWidth, draw=\SimPlotsGridColor, solid},
legend style={at={(1,.1)}, anchor=south east, legend cell align=center, draw=white!15!black}
]
\addplot [color=black, \SimPlotsALineStyle, line width=\SimPlotsLineWidth, mark size=\SimPlotsAMarkSize, mark=\SimPlotsAMarker, mark options={solid, rotate=\SimPlotsARotate, black}]
  table[row sep=crcr]{%
0	0.120198371164457\\
1	0.12168019648405\\
2	0.224425283955936\\
3	0.354140844094292\\
4	0.460152809769026\\
5	0.60717434880495\\
6	0.740306354540847\\
7	0.872216004487267\\
8	1.09471195336698\\
};
\addlegendentry{\SimPlotsLegendCaseA}


\addplot [color=black, \SimPlotsCLineStyle, line width=\SimPlotsLineWidth, mark size=\SimPlotsCMarkSize, mark=\SimPlotsCMarker, mark options={solid, black}]
  table[row sep=crcr]{%
0	0.0385206433825948\\
1	0.210255999207546\\
2	0.524190487680084\\
3	0.947306403110101\\
4	1.42780108880339\\
5	1.95510328486322\\
6	2.45093119197231\\
7	3.01195783075416\\
8	3.65330470585178\\
};
\addlegendentry{\SimPlotsLegendCaseC}

\addplot [color=black, \SimPlotsDLineStyle, line width=\SimPlotsLineWidth, mark size=\SimPlotsDMarkSize, mark=\SimPlotsDMarker, mark options={solid, black}]
  table[row sep=crcr]{%
0	0.117868319514144\\
1	0.160110126022099\\
2	0.408092358343082\\
3	0.752115877657211\\
4	1.15864627688304\\
5	1.61894922872675\\
6	2.04087035642929\\
7	2.51795338618882\\
8	3.1146237495527\\
};
\addlegendentry{\SimPlotsLegendCaseD}

\end{axis}
\end{tikzpicture}%
} }
\subfloat[]{\centering\label{fig:RandomSamplingRMSE_vs_d_Pos}
\resizebox{.48\columnwidth}{!}{
%
%
%
\begin{tikzpicture}

\begin{axis}[%
width=\SimPlotsInnerWidth,
height=\SimPlotsInnerHeight,
at={(0,0)},
scale only axis,
xmin=0,
xmax=8,
xtick={0, 1, 2, 3, 4, 5, 6, 7, 8},
xlabel style={font=\color{white!15!black}},
xlabel={inter-node distance $d$ [m]},
ymin=0,
ymax=\SimPlotsYMaxPos,
ytick distance=.2,
ylabel style={font=\color{white!15!black}},
ylabel={position RMSE [m]},
axis background/.style={fill=white},
axis x line*=bottom,
axis y line*=left,
xmajorgrids=\SimPlotsGrid,
ymajorgrids=\SimPlotsGrid,
grid style={line width=\SimPlotsGridLineWidth, draw=\SimPlotsGridColor, solid},
legend style={at={(0.15,1)}, anchor=north west, legend cell align=center, draw=white!15!black}
]
\addplot [color=black, \SimPlotsELineStyle, line width=\SimPlotsLineWidth, mark size=\SimPlotsEMarkSize, mark=\SimPlotsEMarker, mark options={solid, black}]
  table[row sep=crcr]{%
0	0.063343968140528\\
1	0.063068715005811\\
2	0.0604347292188007\\
3	0.0596021370207057\\
4	0.0632951371415703\\
5	0.0615763230817609\\
6	0.0625297282066808\\
7	0.0619407501809887\\
8	0.060718119496882\\
};
\addlegendentry{\SimPlotsLegendCaseE}

\addplot [color=black, \SimPlotsGLineStyle, line width=\SimPlotsLineWidth, mark size=\SimPlotsGMarkSize, mark=\SimPlotsGMarker, mark options={solid, black}]
  table[row sep=crcr]{%
0	0.063343968140528\\
1	0.063068715005811\\
2	0.0604347292188007\\
3	0.0810796593039542\\
4	0.141146906689986\\
5	0.295305474341817\\
6	0.623139447187105\\
7	1.37423544398855\\
8	5.66830538304392\\
};
\addlegendentry{\SimPlotsLegendCaseG}

\addplot [color=black, \SimPlotsFLineStyle, line width=\SimPlotsLineWidth, mark size=\SimPlotsFMarkSize, mark=\SimPlotsFMarker, mark options={solid, black}]
  table[row sep=crcr]{%
0	0.0633439975816476\\
1	0.0637053371875029\\
2	0.0769631721948384\\
3	0.130533556543416\\
4	0.202620216988131\\
5	0.323306559083692\\
6	0.46727482369433\\
7	0.662066550651092\\
8	0.824734348991316\\
};
\addlegendentry{\SimPlotsLegendCaseF}

\addplot [color=black, \SimPlotsHLineStyle, line width=\SimPlotsLineWidth, mark size=\SimPlotsHMarkSize, mark=\SimPlotsHMarker, mark options={solid, black}]
  table[row sep=crcr]{%
0	0.0310491762240631\\
1	0.0309576916084057\\
2	0.0300903499216143\\
3	0.0288997570161544\\
4	0.0308180104812162\\
5	0.0304470020481956\\
6	0.0306841687925283\\
7	0.0320769387978224\\
8	0.0311383052414917\\
};
\addlegendentry{\SimPlotsLegendCaseH}

\addplot [color=black, \SimPlotsILineStyle, line width=\SimPlotsLineWidth, mark size=\SimPlotsIMarkSize, mark=\SimPlotsIMarker, mark options={solid, black}]
  table[row sep=crcr]{%
0	0.0310491762240631\\
1	0.0309576916084057\\
2	0.0300903499216143\\
3	0.0655487902843982\\
4	0.118380542254134\\
5	0.380379101236687\\
6	0.511100699693482\\
7	1.1223287784039\\
8	1.43918436182665\\
};
\addlegendentry{\SimPlotsLegendCaseI}

\end{axis}
\end{tikzpicture}%
} }
\\[1.5mm]
\subfloat[]{\centering\label{fig:RandomSamplingRMSE_vs_sigmaAngular}
\centering
\resizebox{.75\columnwidth}{!}{
%
%
%
\begin{tikzpicture}

\begin{axis}[%
width=.85\columnwidth,
height=\SimPlotsInnerHeight,
at={(0,0)},
scale only axis,
xmin=0,
xmax=24,
xtick distance=3,
xlabel style={font=\color{white!15!black}},
xlabel={direction error std. dev. $\sigma_{\mathrm{dir}}$ [degree]},
ymin=0,
ymax=\SimPlotsYMaxPos,
ytick distance=.2,
ylabel style={font=\color{white!15!black}},
ylabel={position RMSE [m]},
axis background/.style={fill=white},
axis x line*=bottom,
axis y line*=left,
xmajorgrids=\SimPlotsGrid,
ymajorgrids=\SimPlotsGrid,
grid style={line width=\SimPlotsGridLineWidth, draw=\SimPlotsGridColor, solid},
legend style={at={(.465,1)}, anchor=north east, legend cell align=center, draw=white!15!black}
]
\addplot [color=black, \SimPlotsELineStyle, line width=\SimPlotsLineWidth, mark size=\SimPlotsEMarkSize, mark=\SimPlotsEMarker, mark options={solid, black}]
  table[row sep=crcr]{%
0	0.061425436429748\\
2	0.0696521036736242\\
4	0.0864506509494685\\
6	0.107426578318944\\
8	0.137705573639308\\
10	0.162532296028431\\
12	0.189728015118519\\
14	0.222749893779566\\
16	0.25362947667072\\
18	0.283285033427662\\
20	0.320704342964493\\
22	0.355004016019742\\
24	0.391376767300882\\
};
\addlegendentry{\SimPlotsLegendCaseE}

\addplot [color=black, \SimPlotsGLineStyle, line width=\SimPlotsLineWidth, mark size=\SimPlotsGMarkSize, mark=\SimPlotsGMarker, mark options={solid, black}]
  table[row sep=crcr]{%
0	0.0614254364297481\\
2	0.0696521036736242\\
4	0.0865683824360939\\
6	0.111854593229833\\
8	0.162569799261957\\
10	0.306964918137443\\
12	0.391332805446333\\
14	0.463306690833742\\
16	0.634728415834642\\
18	1.19927073536325\\
20	2.29604037874681\\
22	177.440604272851\\
24	5.04023390801859\\
};
\addlegendentry{\SimPlotsLegendCaseG}

\addplot [color=black, \SimPlotsFLineStyle, line width=\SimPlotsLineWidth, mark size=\SimPlotsFMarkSize, mark=\SimPlotsFMarker, mark options={solid, black}]
  table[row sep=crcr]{%
0	0.0793677844303229\\
2	0.0893653058999901\\
4	0.113990257058639\\
6	0.145465843957094\\
8	0.190878412645562\\
10	0.225414541178721\\
12	0.257984297484831\\
14	0.305009077320069\\
16	0.349829357797872\\
18	0.38843906447164\\
20	0.434861474438561\\
22	0.46500312375525\\
24	0.515390165050095\\
};
\addlegendentry{\SimPlotsLegendCaseF}

\addplot [color=black, \SimPlotsHLineStyle, line width=\SimPlotsLineWidth, mark size=\SimPlotsHMarkSize, mark=\SimPlotsHMarker, mark options={solid, black}]
  table[row sep=crcr]{%
0	0.0317542145247407\\
2	0.781111620846505\\
4	1.18625218158732\\
6	1.34246545982128\\
8	1.46589688002535\\
10	1.50825454321524\\
12	1.5712727017512\\
14	1.62977052319567\\
16	1.67474669583841\\
18	1.73136249607892\\
20	1.79436076336559\\
22	1.8766355129089\\
24	1.95277948864262\\
};
\addlegendentry{\SimPlotsLegendCaseH}

\addplot [color=black, \SimPlotsILineStyle, line width=\SimPlotsLineWidth, mark size=\SimPlotsIMarkSize, mark=\SimPlotsIMarker, mark options={solid, black}]
  table[row sep=crcr]{%
0	0.0317542145247406\\
2	0.781111620846505\\
4	1.18654654234302\\
6	1.34149568065242\\
8	1.45657616670616\\
10	1.48402871167481\\
12	1.52692266692572\\
14	1.55761273831397\\
16	1.57825113145008\\
18	1.6881829007725\\
20	1.77748840164659\\
22	4.65472726892709\\
24	2.24919318892097\\
};
\addlegendentry{\SimPlotsLegendCaseI}

\end{axis}
\end{tikzpicture}%
}
}
\caption{Numerical evaluation of the estimation RMSE versus different parameters. The experiment assumes $\NObs = 3$ observers, number of MPCs $K_1 = K_2 = K_3 = 4 \ \Rightarrow \ K = 12$, distance $d = 2\unit{m}$, $\sigma = 0.2\unit{ns}$.}
\label{fig:RandomSamplingRMSE_d2m}
\end{figure}
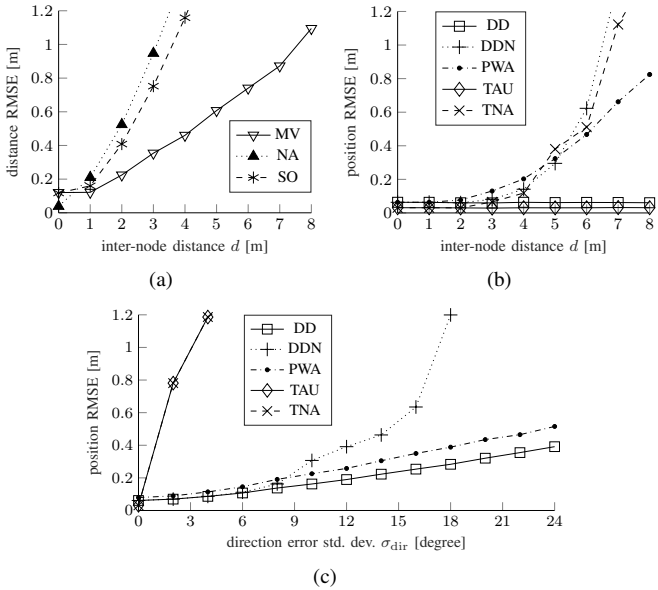

\begin{figure}
\centering
\subfloat[]{\centering\label{fig:RandomSamplingRMSEvsK_d2m_Dist}
\resizebox{.48\columnwidth}{!}{
%
%
%
\begin{tikzpicture}

\begin{axis}[%
width=\SimPlotsInnerWidth,
height=\SimPlotsInnerHeight,
at={(0,0)},
scale only axis,
xmin=2,
xmax=8,
xtick={2, 3, 4, 5, 6, 7, 8},
xlabel style={font=\color{white!15!black}},
xlabel={number of MPCs $K$},
ymin=0,
ymax=\SimPlotsYMaxDist,
ytick distance=.2,
ylabel style={font=\color{white!15!black}},
ylabel={distance RMSE [m]},
axis background/.style={fill=white},
axis x line*=bottom,
axis y line*=left,
xmajorgrids=\SimPlotsGrid,
ymajorgrids=\SimPlotsGrid,
grid style={line width=\SimPlotsGridLineWidth, draw=\SimPlotsGridColor, solid},
legend style={at={(.07,.07)}, anchor=south west, legend cell align=center, draw=white!15!black}
]
\addplot [color=black, \SimPlotsALineStyle, line width=\SimPlotsLineWidth, mark size=\SimPlotsAMarkSize, mark=\SimPlotsAMarker, mark options={solid, rotate=\SimPlotsARotate, black}]
  table[row sep=crcr]{%
2	1.39850363503541\\
3	0.919538408629156\\
4	0.663488658937486\\
5	0.540433505920266\\
6	0.454998393985327\\
7	0.400140854802128\\
8	0.334546087853132\\
};
\addlegendentry{\SimPlotsLegendCaseA}


\addplot [color=black, \SimPlotsCLineStyle, line width=\SimPlotsLineWidth, mark size=\SimPlotsCMarkSize, mark=\SimPlotsCMarker, mark options={solid, black}]
  table[row sep=crcr]{%
2	1.40176258864066\\
3	1.14567080983269\\
4	0.965471223276855\\
5	0.830974089767105\\
6	0.756776800173488\\
7	0.687052823507233\\
8	0.621155682131891\\
};
\addlegendentry{\SimPlotsLegendCaseC}

\addplot [color=black, \SimPlotsDLineStyle, line width=\SimPlotsLineWidth, mark size=\SimPlotsDMarkSize, mark=\SimPlotsDMarker, mark options={solid, black}]
  table[row sep=crcr]{%
2	1.34334007126534\\
3	0.940996227783508\\
4	0.734214982697216\\
5	0.639494923472891\\
6	0.585832077951637\\
7	0.541778061671576\\
8	0.496306308188996\\
};
\addlegendentry{\SimPlotsLegendCaseD}

\end{axis}
\end{tikzpicture}%
} }
\subfloat[]{\centering\label{fig:RandomSamplingRMSEvsK_d2m_Pos}
\resizebox{.48\columnwidth}{!}{
%
%
%
\begin{tikzpicture}

\begin{axis}[%
width=\SimPlotsInnerWidth,
height=\SimPlotsInnerHeight,
at={(0,0)},
scale only axis,
unbounded coords=jump,
xmin=3,
xmax=8,
xtick={2, 3, 4, 5, 6, 7, 8},
xlabel style={font=\color{white!15!black}},
xlabel={number of MPCs $K$},
ymin=0,
ymax=\SimPlotsYMaxPos,
ytick distance=.2,
ylabel style={font=\color{white!15!black}},
ylabel={position RMSE [m]},
axis background/.style={fill=white},
axis x line*=bottom,
axis y line*=left,
xmajorgrids=\SimPlotsGrid,
ymajorgrids=\SimPlotsGrid,
grid style={line width=\SimPlotsGridLineWidth, draw=\SimPlotsGridColor, solid},
legend style={at={(.9,.9)}, anchor=north east, legend cell align=center, draw=white!15!black}
]
\addplot [color=black, \SimPlotsELineStyle, line width=\SimPlotsLineWidth, mark size=\SimPlotsEMarkSize, mark=\SimPlotsEMarker, mark options={solid, black}]
  table[row sep=crcr]{%
4	1.00460691657079\\
5	0.375180533039729\\
6	0.156306755702613\\
7	0.109100063388235\\
8	0.0912304803167047\\
};
\addlegendentry{\SimPlotsLegendCaseE}


\addplot [color=black, \SimPlotsFLineStyle, line width=\SimPlotsLineWidth, mark size=\SimPlotsFMarkSize, mark=\SimPlotsFMarker, mark options={solid, black}]
  table[row sep=crcr]{%
4	1.10387268373946\\
5	0.405864510211493\\
6	0.197617157195956\\
7	0.141692743123191\\
8	0.11700383625673\\
};
\addlegendentry{\SimPlotsLegendCaseF}

\addplot [color=black, \SimPlotsHLineStyle, line width=\SimPlotsLineWidth, mark size=\SimPlotsHMarkSize, mark=\SimPlotsHMarker, mark options={solid, black}]
  table[row sep=crcr]{%
5	0.0551158511509493\\
6	0.0472844195763638\\
7	0.0432577179250979\\
8	0.0401039675637527\\
};
\addlegendentry{\SimPlotsLegendCaseH}


\end{axis}
\end{tikzpicture}%
} }
\caption{Numerical evaluation of the estimation RMSE versus the number of MPCs $K$ for $\NObs = 1$ observers, $K_1 = K$, $d = 2\unit{m}$, $\sigma = 0.2\unit{ns}$.}
\label{fig:RandomSamplingRMSEvsK_d2m}
\end{figure}
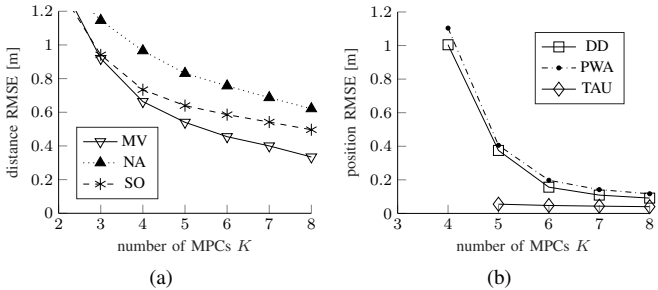

\Cref{fig:RandomSamplingRMSE_vs_sigmaAngular} shows a very important difference between the two different approaches for position estimation. In particular, the $\delayDiff[]$-based position LSE (cases \SimPlotsLegendCaseE{}, \SimPlotsLegendCaseF{}, \SimPlotsLegendCaseG{}) copes with erroneous MPC direction measurements very well. The $\tau$-based position LSE (cases \SimPlotsLegendCaseH{}, \SimPlotsLegendCaseI{}) however deteriorates even for small directional errors. This is caused by the terms $\dirVectB[k o]\!-\dirVectA[k o]$ in the matrix ${\bf G}$ when computing $\d\LSE\AssViaTau$ in \Cref{eq:EstimateRelLocViaTauDetails}. In simpler terms, it is clear that the underlying property $\d = \cd\delayTrueB[k o] \dirVectB[k o] - \cd\delayTrueA[k o] \dirVectA[k o]$
from \Cref{eq:VectorEquality} can only lead to an accurate estimate of $\d$ if $\dirVectA[k o]$ and $\dirVectB[k o]$ are measured accurately.

The complimentary experiment in \Cref{fig:RandomSamplingRMSEvsK_d2m} studies the effect of the number of MPCs $K$ on the estimation accuracy. It considers only one observer ($\NObs=1$). Clearly, all estimators benefit from an increasing $K$. For $\d\LSE\AssViaDiff$ (\SimPlotsLegendCaseE{}, \SimPlotsLegendCaseF{}) it seems particularly fruitful to exceed the minimum $K$ of $4$ by some margin, to ensure that $\E\E\Tr$ in \Cref{eq:displLSEAssViaDiff} is well-conditioned.

	\section{Summary \& Outlook}
	\label{sec:Summary}
	For a recently proposed paradigm for UWB wireless localization, we derived novel distance and position estimators for various cases.
Future work shall embed them in localization algorithms (e.g, for cooperative network localization or proximity detection) together with robust MPC selection schemes,
apply temporal filtering, and conduct practical field trials
	
	
	
		    

%


	
\bibliographystyle{IEEEtran}
\bibliography{IEEEabrv,ref}
	
	
\pdfinfo{
	   /Author(Gregor Dumphart, Robin Kramer, Armin Wittneben)
	   /Title(\OurPaperTitle)
	   /Subject(Wireless Ranging and Localization)
	   /Keywords(\OurIndexTerms{})
}

\end{document}